\documentclass[bibyear]{aa}

\usepackage{graphicx}
\usepackage{txfonts}
\usepackage{amsmath}	
\usepackage{ulem}
\usepackage{multirow}
\usepackage{graphicx}
\usepackage{caption}
\usepackage{subcaption}
\usepackage{natbib}
\usepackage{amssymb}	
\usepackage{xcolor}
\usepackage{orcidlink}

\newcommand\Msun{\mathrm{M}_\odot}
\newcommand\Zsun{\mathrm{Z}_\odot}
\newcommand\Mstar{\mathcal{M}_\star}

\newcommand\cmc{\texttt{CMC}}
\newcommand\firebox{\texttt{FIREbox}}

\makeatletter
\renewcommand*\aa@pageof{, page \thepage{} of \pageref*{LastPage}}
\makeatother

\defcitealias{grudic22}{Paper~I}
\defcitealias{rodriguez23}{Paper~II}
\defcitealias{bruel24}{Paper~III}

\begin{document}

   \title{Great Balls of FIRE\\ IV. The contribution of massive star clusters to the astrophysical population of merging binary black holes}

   \subtitle{}

   \author{T. Bruel\inst{1,2,3}\fnmsep\thanks{e-mail: tristan.bruel@oca.eu}\orcidlink{0000-0002-1789-7876}
          \and
          A. Lamberts\inst{3,4}\orcidlink{0000-0001-8740-0127}
          \and
          C. L. Rodriguez\inst{5}\orcidlink{0000-0003-4175-8881}
          \and
          R. Feldmann\inst{6}\orcidlink{0000-0002-1109-1919}
          \and
          M. Y. Grudi\'c\inst{7}\thanks{NASA Hubble Fellow}\orcidlink{0000-0002-1655-5604}
          \and
          J. Moreno\inst{8,9}\orcidlink{0000-0002-3430-3232}
          }

   \institute{
            Dipartimento di Fisica ``G. Occhialini'', Universit\`a degli Studi di Milano-Bicocca, Piazza della Scienza 3, 20126 Milano, Italy
            \and 
            INFN, Sezione di Milano-Bicocca, Piazza della Scienza 3, 20126 Milano, Italy
            \and
            Laboratoire Lagrange, Universit\'e C\^ote d'Azur, Observatoire de la C\^ote d'Azur, CNRS, Bd de l’Observatoire, 06300, France
            \and
            Laboratoire Artemis, Universit\'e C\^ote d'Azur, Observatoire de la  C\^ote d'Azur, CNRS, Bd de l’Observatoire, 06300, France
            \and
            Department of Physics and Astronomy, University of North Carolina at Chapel Hill, 120 E. Cameron Ave, Chapel Hill, NC, 27599, USA
            \and
            Institute for Computational Science, University of Zurich, Winterthurerstrasse 190, 8057 Zurich, Switzerland
            \and
            Center for Computational Astrophysics, Flatiron Institute, 162 5th Ave, New York, NY 10010, USA
            \and
            Department of Physics and Astronomy, Pomona College, Claremont, CA 91711, USA
            \and
            Carnegie Observatories, Pasadena, CA 91101, USA
            }

   \date{}

  \abstract
   {The detection of over a hundred gravitational wave signals from double compacts objects, reported by the LIGO-Virgo-KAGRA Collaboration, have confirmed the existence of such binaries with tight orbits. Two main formation channels are generally considered to explain the formation of these merging binary black holes (BBHs): the isolated evolution of stellar binaries, and the dynamical assembly in dense environments, namely star clusters. Although their relative contributions remain unclear, several analyses indicate that the detected BBH mergers probably originate from a mixture of these two distinct scenarios.}
   {We study the formation of massive star clusters across time and at a cosmological scale to estimate the contribution of these dense stellar structures to the overall population of BBH mergers.}
   {To this end, we propose three different models of massive star cluster formation based on results obtained with zoom-in simulations of individual galaxies. We apply these models to a large sample of realistic galaxies identified in the $(22.1\ \mathrm{Mpc})^3$ cosmological volume simulation \firebox. Each galaxy in this simulation has a unique star formation rate, with its own history of halo mergers and metallicity evolution. Combined with predictions obtained with the Cluster Monte Carlo code for stellar dynamics, we are able to estimate populations of dynamically formed BBHs in a collection of realistic galaxies.}
   {
   Across our three models, we infer a local merger rate of BBHs formed in massive star clusters consistently in the range $1-10\ \mathrm{Gpc}^{-3}\mathrm{yr}^{-1}$. 
   Compared with the local BBH merger rate inferred by the LIGO-Virgo-KAGRA Collaboration (in the range $17.9-44\ \mathrm{Gpc}^{-3}\mathrm{yr}^{-1}$ at $z=0.2$), this could potentially represent up to half of all BBH mergers in the nearby Universe.
   This shows the importance of this formation channel in the astrophysical production of merging BBHs.
   We find that these events preferentially take place in the most massive galaxies.}
   {}


   \maketitle
%

\section{Introduction}

The ever-growing list of gravitational wave signals detected by the LIGO-Virgo-KAGRA Collaboration (LVK)\footnote{Find LVK O4 public alerts at \url{https://gracedb.ligo.org/superevents/public/O4/}} has now definitively proved the existence of double compact objects, and in particular of binary black holes (BBHs), with initial orbital separations small enough for the two objects to merge in less than a Hubble time \citep[see the third Gravitational Wave Transient Catalog GWTC-3,][]{gwtc3}. However, the astrophysical processes leading to the formation of these binaries remain challenging to constrain. Recent analyses of the physical properties of these detected BBHs show growing evidence that several formation channels are likely to be involved \citep[see e.g.][]{zevin21,arcasedda23,ray24}.

As massive stars are expected to end their lives as black holes and are observed to exist primarily in binaries or multiples \citep[see e.g.][]{sana12, moe17}, a natural scenario to explain the formation of these BBHs seems to be the conjoint evolution of two massive stars. This formation channel has been extensively studied over the past decades \citep[e.g.][]{bethe98, belczynski02, dominik12, belczynski16, eldridge16, stevenson17, giacobbo18, neijssel19, santoliquido21, tauris23} and significant progress has been made in our understanding of binary stellar evolution. Most studies rely on population synthesis techniques that make it possible to numerically build large populations of double compact objects and, using models of metallicity-dependent star formation rate to describe the evolution of the Universe, to distribute their mergers over cosmic time. However, a number of uncertainties still persist in this framework, both in the description of the Universe and in the prescriptions used to evolve binary stars \citep[see e.g.][for studies on the relative importance of each of these two aspects]{broekgaarden22, santoliquido22, srinivasan23}.

A second possible pathway that could lead to the formation of two BHs orbiting each other with very close separation is to consider dynamical interactions in dense environments. In star clusters, where the presence of stellar BHs has been confirmed both through X-ray and radio observations \citep[see e.g.][]{maccarone07,strader12,chomiuk13,miller15,shishkovsky18} and radial velocity measurements of detached star-BH binaries \citep[see e.g.][]{giesers18, giesers19}, mass segregation and three-body interactions are expected to enhance the pairing of BHs in dynamically-hard binaries \citep[see e.g.][]{morscher15}. These dense stellar structures, such as young clusters, globular clusters or even nuclear clusters, constitute environments favourable to the formation of GW sources in the form of tight BBHs \citep[see e.g.][]{sigurdsson93, zwart00, rodriguez15, rodriguez16, banerjee17, diCarlo20}. As both direct and Monte Carlo simulations of dynamics in star clusters can be numerically quite expensive, most studies consider a rather modest number of simulated star clusters in a predetermined parameter space. The recent development of semi-analytical codes for the rapid evolution of star clusters, such as \texttt{FASTCLUSTER} \citep{mapelli21} or \texttt{RAPSTER} \citep{kritos24}, now makes it possible to operate BBH population synthesis for much larger numbers of star clusters with a wide range of physical properties. The trends observed in the results of their evolution can then be used to predict cosmologically statistically significant quantities, such as the evolution of the BBH merger rate in star clusters over cosmic time \citep{rodriguez18, kremer20, mapelli22, ye24}.

Although stellar populations are in most cases described with the combination of well-known models, such as a star formation rate \citep[SFR,][]{madau14} and an initial mass function \citep{kroupa01}, it is much more complex to describe the evolution of star cluster physical properties over cosmic time \citep[see e.g.][for a review]{krumholz19}. In particular, several quantities used to describe the formation of star clusters have been found to vary depending on the galactic environment, including the cluster formation efficiency \citep{ginsburg18}, the cluster initial mass function with a potential high-mass truncation \citep{wainer22}, and the mass-radius relation \citep{brown21}.
In addition, the evolution of a single star cluster over time is virtually impossible to predict using simple models alone, and shows major changes between its initial state and what we can observe as its present-day aspect. Both the formation and the evolution of star clusters are also highly dependent on a number of environmental effects \citep[see e.g.][]{rossi16,suin22}.
This implies that a realistic model of the formation and evolution of star clusters over cosmic time should take into account the evolution of the environmental conditions specific to each galaxy, and this with a sufficiently large number of different galaxies to be able to represent their diversity in the Universe.
Here again, a number of uncertainties remain as to the physical properties of star clusters used to describe a complete cosmological population, and as to the diversity of results predicted from cluster dynamics for different initial conditions.

A second approach to study the cosmological population of dynamically formed BBHs is to use a large volume cosmological simulation as the natural environment in which stars and clusters can form. Given the high computational cost of this type of simulation, a compromise must always be found between the spatial resolution of the simulation, its volume, and the redshift to which it evolves. To date, the majority of numerical simulations with a sub-parsec scale resolution necessary to resolve the formation of massive star clusters have only been made in small volumes and/or at large redshifts \citep[see e.g.][the latter resolving the formation of star clusters in zoom-in simulations of cosmological volume of $5\mathrm{cMpc}^3$ from $z=100$ down to $z=10.5$]{boley09, kimm16, lahen19, calura22, calura24}.
In order to predict the formation rate and physical properties of the star clusters that would populate any given simulated galaxy up to the present-day, semi-analytical models are then required. Studies of this kind have already been carried out, with favourable results when compared with the population of globular clusters (GCs) observed in local galaxies \cite[e.g.][]{li14, pfeffer18, grudic21}.

This paper follows the Great Balls of FIRE series \citep[][hereafter \citetalias{grudic22}, \citetalias{rodriguez23}, and \citetalias{bruel24} respectively]{grudic22, rodriguez23, bruel24}. Here we elaborate upon these analyses of the formation and evolution of massive star clusters inside cosmological zoom-in simulations of galaxies to develop models of cluster formation that can be applied to a larger cosmological simulation containing a wide variety of simulated galaxies.
We consider here only the production of BBHs in massive star clusters, defined as the star clusters with initial masses larger than $6\times10^4\ \Msun$ (value selected to maintain consistency with our previous studies \citetalias{rodriguez23} and \citetalias{bruel24}, see in \S\ref{sub:models} below for more details). Making use of the large catalogue of massive clusters already integrated forward in time in \citetalias{rodriguez23,bruel24} with the Cluster Monte Carlo code \citep[\cmc][]{pattabiraman13, rodriguez22}, we are able to make predictions about the contribution of different galaxies and environments to the dynamical population of merging BBHs. Throughout this study and in all that follows, we only consider BBHs that have already merged by $z=0$ and every reference to the term `BBHs' should be interpreted as `merging BBHs'.

After describing the cosmological volume simulation that we use as a large sample of realistic galaxies and the three models of cluster formation that we consider (Section~\ref{sec:method}), we present our predictions on massive star clusters in galaxies at a cosmological scale and on the BBH mergers that these clusters produce (Section~\ref{sec:results}).
We discuss implications of our findings and outline future extensions of this work (Section~\ref{sec:discussion}) and conclude on the significance of our results (Section~\ref{sec:conclusions}).

\section{Method}
\label{sec:method}

In this work, we aim to estimate the physical properties and merger rate of dynamically formed BBHs at a cosmological scale. We use a large sample of galaxies identified in a cosmological volume simulation (\S\ref{sub:FIREbox}) and propose three different semi-analytic models for massive star cluster formation (\S\ref{sub:models}) to model the population of star clusters in each galaxy. We build upon the large collection of massive star clusters (1500) already integrated with the code \cmc\ in \citetalias{rodriguez23,bruel24} to develop a grid-matching algorithm that makes it possible to estimate the BBH mergers formed dynamically in any population of massive star clusters (\S\ref{sub:grid}).

\subsection{Realistic star formation histories from individual galaxies in \firebox}
\label{sub:FIREbox}

\begin{figure*}
    \centering
    \includegraphics{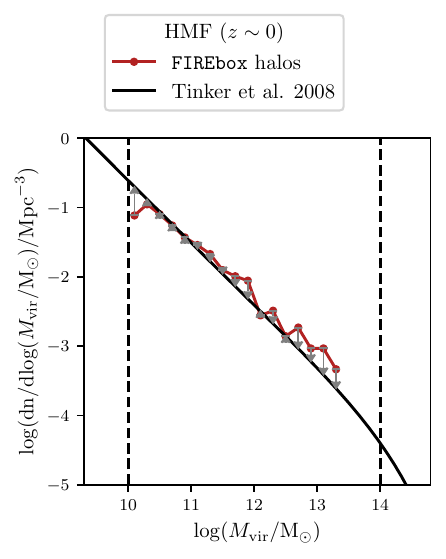}
    \hspace{1em}
    \includegraphics{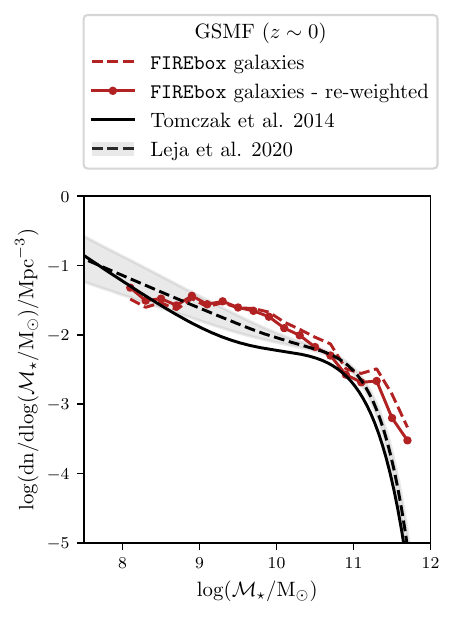}
    \caption{\textbf{Left:} Halo mass function (HMF) of the 979 galaxies identified inside \firebox\ within the mass range $10\leq\mathrm{log}(M_\mathrm{vir}/\Msun)\leq14$. Vertical dashed lines show the range of the virial masses used for halo identification. Grey arrows indicate the re-weighting operated to match the HMF in \firebox\ with a `universal' HMF extracted from dark matter cosmological simulation \citep[in black line, from][]{tinker08}. \textbf{Right:} Galaxy stellar mass function (GSMF) for this same set of galaxies identified inside \firebox, and with each galaxy re-weighted with the operation presented in the left-hand panel. Two examples of GSMF built on observations of galaxies in the local Universe are also shown. The black solid line is the best-fit GSMF at $z\sim 0.1$ from \citet{tomczak14} and the dashed line is the median GSMF generated using the continuity model parameters of \citet{leja20} at $z=0$. The shaded grey area corresponds to the 16-84 \% uncertainties associated to this model.}
    \label{fig:HMF}
\end{figure*}

\firebox\ \citep{feldmann23} is a large scale cosmological simulation created as part of the FIRE project \citep{hopkins14}. It represents a cosmological volume of $(22.1\mathrm{cMpc})^3$ evolved from redshift 120 to the present-day, with 1200 snapshots spaced in time. It uses the same FIRE-2 physics model as the zoom-in simulations used in \citetalias{bruel24}.
FIRE-2 models radiative cooling and heating across the range $10-10^{10}\ \mathrm{K}$, including free-free, photo-ionisation and recombination, Compton, photo-electric and dust collisional, cosmic ray, molecular, metal-line, and fine-structure processes. Photo-ionisation and heating from a redshift-dependent, spatially uniform ultraviolet background \citep{faucher09} is also taken into account. 
Star formation occurs in self-gravitating, self-shielding, and Jeans unstable dense molecular gas ($n > 300\ \mathrm{cm}^{-3}$). All feedback event rates, including energy, momentum, mass and metal injection from type Ia and type II supernovae, and stellar winds luminosities, are tabulated from stellar evolution models \citep[STARBUST99;][]{leitherer99} assuming a \citet{kroupa01} initial stellar mass function (IMF). A sub-grid model is applied for the turbulent diffusion of metals in gas \citep{escala17}.

We choose \firebox\ specifically in the present work to maintain a certain consistency with our previous analysis, but also because it offers a spatial dynamic range of $\sim10^6$, about an order of magnitude higher than most large-volume simulations. This corresponds to a spatial resolution of $\sim20\ \mathrm{pc}$, well suited to explore the internal structure of galaxies, but still too large to resolve individual star clusters. The baryonic mass resolution is $m_\mathrm{b}\sim6.3\times10^4\ \Msun$. 
Although this simulation is considered a large volume and contains thousands of galaxies, the total volume of the simulated cube is not yet sufficient to be representative of our Universe. Indeed, most analyses suggest that the typical homogeneity scale of our Universe is of the order $\mathcal{R}_\mathrm{H}\gtrsim100\ h^{-1}\mathrm{Mpc}$ \citep[see e.g.][]{scrimgeour12, ntelis18}. This is an obvious limitation of the analysis we carry out here and should be borne in mind in all that follows. The effect of the cosmic variance inherent in this restricted volume will be discussed in more detail below.

\begin{figure*}
    \centering
    \includegraphics{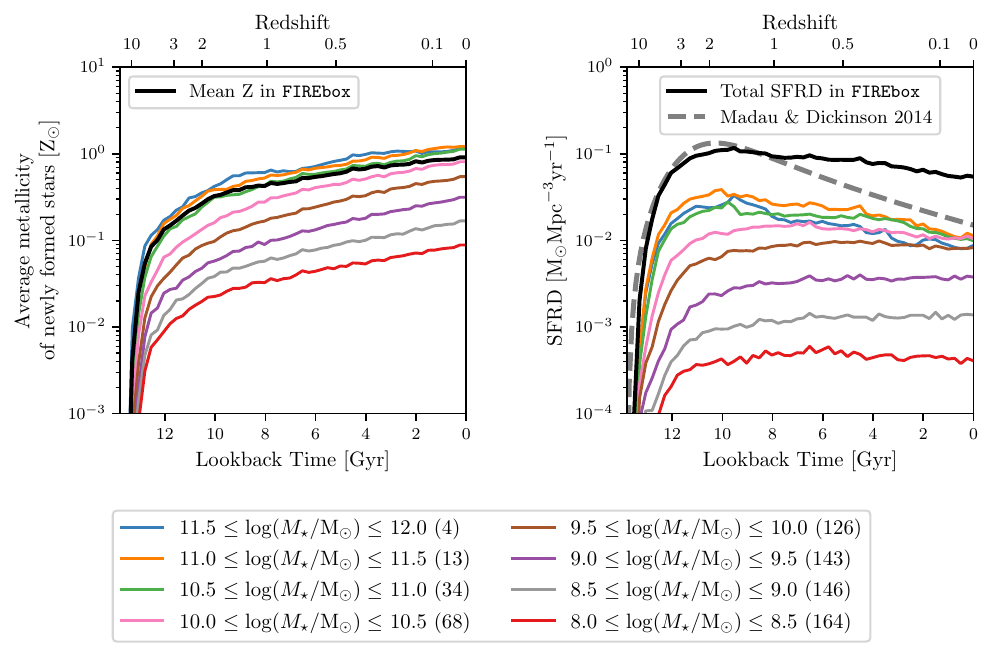}
    \caption{\textbf{Left:} Mass-weighted average of the metallicity of newly formed stars for the 979 galaxies identified inside \firebox. The black line indicates the average metallicity is in the entire volume, and the different colours indicate galaxies grouped in different bins of stellar mass. The number of galaxies in each mass bin is indicated in brackets.
    The contribution of each galaxy is re-weighted as described in \S\ref{sub:FIREbox} (and as pointed out in the left-hand panel of Figure \ref{fig:HMF}). 
    \textbf{Right:} Star formation rate density (SFRD) in the entire volume of \firebox\ (black line) and for galaxies grouped in different bins of stellar mass (coloured lines). 
    As a comparison, the grey dashed line shows the functional form of the cosmic SFRD proposed by \citet{madau14}.}
    \label{fig:SFRs in box}
\end{figure*}

We use the halo structures already identified inside \firebox\ with the AMIGA Halo Finder \citep[AHF,][]{knollmann09}. In particular, we identify in the last snapshot (corresponding to redshift 0) 979 halos with present-day virial masses $M_\mathrm{vir}\geq10^{10}\ \Msun$ (which corresponds to a lower limit on galaxy stellar mass of roughly $\Mstar\geq10^8\ \Msun$). This provides us with a large sample of realistic galaxies that can be used to study the formation of star clusters inside each of them across cosmic time. 
Given the empirical near-constant fraction of the total mass of GCs in a galaxy divided by its total mass (dark and baryonic matter), no massive star clusters are expected to form in less massive dwarf galaxies \citep{harris17,chen23,bruel24}. Using high-resolution optical imaging of quiescent galaxies in the Virgo cluster, \citet{sanchezJanssen19} find that the GC occupation fraction rapidly falls off for galaxy stellar masses below $10^9\Msun$. Thus, our galaxy stellar mass cut-off $\Mstar\geq10^8\ \Msun$ will have only a minor impact on the results presented here on massive cluster formation and dynamically formed BBHs.
For each galaxy we extract the following present-day parameters: the virial radius $R_\mathrm{vir}$, the halo mass $\mathcal{M}_h$, the stellar radius $R_{\star,90}$ (computed as the spherical radius that encloses 90\% of the stellar mass within 20 kpc, see \citet{wetzel23}), and the corresponding stellar mass $M_{\star,90}$. In what follows, the stellar mass $\Mstar$ of a galaxy identified in \firebox\ always corresponds to the computed value $M_{\star,90}$.

We show the halo mass function (HMF) obtained from these halos in the left-hand panel of Figure \ref{fig:HMF}. The HMF presented in \citet{tinker08}, obtained from a large set of collisionless dark-matter only cosmological simulations with flat $\Lambda-\mathrm{CDM}$ cosmology, is also shown as a comparison. 
There appears to be a mismatch between the HMF realised in \firebox\ and this `universal' form of the HMF.
\citet{feldmann23} explain this discrepancy by arguing that the HMF that exists and evolves throughout the cosmological volume of the \firebox\ simulation differs from a true local HMF because of a cosmic variance that particularly affects smaller volumes. This variance is due to the finite size of the box and its limited numerical resolution.

As we are interested in comparing the predictions of this numerical approach with the LVK detections of BBH mergers, this sample of simulated galaxies needs to compare with the actual distribution of galaxies in the Universe. 
A reasonable match at redshift 0 would not be sufficient to guarantee that the properties of our galaxies are realistic at all epochs, but it is already a first step towards constraining the significance of our subsequent analysis.
To account for the cosmic variance described above, we re-weight all the identified halos to ensure that their HMF matches the one of \citet{tinker08} (as indicated by the arrows in the left-hand panel of Figure \ref{fig:HMF}). In practice, we compute the present-day HMF in \firebox\ using the logarithm of the halo masses at redshift $z=0$ and 0.2 wide bins between 10 and 14, and then calculate the fraction $\mathrm{HMF}_\mathrm{FIREbox} /\mathrm{HMF}_\mathrm{Tinker}$ for each bin \citep[see also][for a more complete and detailed version of such re-weighting process]{feldmann23}. These values are saved as weights for all of the 979 galaxies.
This approach is equivalent to assuming that the halos and galaxies in \firebox\ may very well have a realistic evolution, but are not in the right number to represent a global realistic population and their abundance (or number density) needs to be modified accordingly.

The galaxy stellar mass function (GSMF) obtained from these galaxies is shown in the right-hand panel of Figure \ref{fig:HMF}. As a comparison, the best-fit GSMF presented in \citet{tomczak14} at redshift $z\sim0.1$ is plotted in the same Figure. 
Furthermore, we also provide an estimate of the GSMF at $z=0$ built using the non-parametric modeling of \citet{leja20}. With the exception of an excess of both the $\lesssim 10^{10}\Msun$ and the highest mass galaxies, the GSMF in \firebox\ is in qualitative agreement with the latter estimate. 

For each identified galaxy, we extract the formation time and metallicity (computed here as the mass fraction of all metals tracked in the simulation) for all the star particles that lie, at the present-day, inside the sphere of stellar radius $R_{\star,90}$. In the left-hand panel of Figure \ref{fig:SFRs in box}, we show the evolution of the average metallicity of newly formed stars as a function of time. The mean metallicity follows an overall similar trend when considering the entire simulation and when grouping galaxies by their present-day stellar masses. We refer the reader to \citet{bassini24} for a more detailed analysis of the mass-metallicity relation in the \firebox\ simulation.
Consistently with observations of the mass-metallicity relation \citep[see e.g.][]{mannucci10,nakajima23}, the most massive galaxies are also the most-metal rich.

For all star particles associated with identified galaxies, we use their formation time, metallicity, and mass at $z=0$ to calculate their initial masses. Thus we gain access to the mass-weighted formation history of this population of stars. This calculation yields an `archaeological' star formation history for all the \firebox\ galaxies. The resulting cosmic SFRD is shown in the right-hand panel of Figure \ref{fig:SFRs in box}. It is in good agreement with the observed cosmic SFRD \citep{madau14} up to redshifts $z\geq1$, but over-predicts the SFRD at lower redshifts \citep[see also][]{feldmann23}. Even with the re-weighting operation to account for the cosmic variance in this type of simulation, there is still an excess of $\sim0.6$ dex in the total SFRD at present-day ($z=0$). This excess of star formation is mostly driven by a few very massive galaxies (17 galaxies with stellar masses in the range $[10^{11},10^{12}]\ \Msun$).

It is important to note that, in this large scale cosmological simulation, no prescriptions for supermassive BHs and AGN feedback are included. As a result, the fraction of quiescent massive galaxies is significantly under-predicted in \firebox\ and the contribution of these massive galaxies to the cosmic SFRD at low redshifts is over-estimated \citep{feldmann23}. Furthermore, despite their overall negative impact on star formation and galaxy growth, recent simulations have shown that quasar winds could potentially induce local positive feedback on both star formation \citep{mercedes23} and star clusters \citep{mercedes24}.

Since the overall excess of star formation in \firebox\ occurs mainly in very massive galaxies and at low redshifts, it involves the formation of metal-rich stars and star clusters, which can be expected to make only a small contribution to the total population of merging BBHs \citep{diCarlo20,bruel24}.
Indeed, massive stars formed in metal-rich environments experience higher mass loss through stellar winds \citep[see e.g.][]{vink01}, which means that a smaller fraction of them will have a final mass high enough to collapse into a BH. Furthermore, for massive stars in binary systems (whether isolated or inside a star cluster), mass loss produces a widening of their orbits, which results in a higher fraction of BBHs that will merge in more than a Hubble time.
In practice, we find that $\sim98\%$ of all the star particles formed at redshifts $z\leq1$ in galaxies with present-day stellar masses $\Mstar\geq10^{11}\ \Msun$ have values of metallicity higher than $\Zsun$. Such high metallicity values tend to prevent the production of merging BBHs, even in the most massive star clusters (see Figure \ref{fig:grid} further on).
For this reason, it appears reasonable to assume that the lack of quiescent galaxies in \firebox\ will not have a major impact on further analysis. 

To account for the quenching of star formation due to additional physical processes not implemented in \firebox, \citet{feldmann23} propose to re-weight all galaxies by a quenching factor $1-f_\mathrm{Q}$, using $f_\mathrm{Q}=20\%$, $45\%$, and $90\%$ for galaxies with stellar masses in the ranges $[10^{9},10^{10}]\ \Msun$, $[10^{10},10^{11}]\ \Msun$, and $\geq10^{11}\ \Msun$ respectively \citep[values taken from][]{behroozi19}. This operation brings the predictions for the SFRD in the simulation at the lowest redshifts in much better agreement with observations \citep{feldmann23}. However, as the evolution of BBHs through the emission of GWs can be an extremely slow process, a number of their mergers observed in the local Universe can very well originate from stellar binaries or star clusters formed at much higher redshifts \citep[see e.g.][]{fishbach21}. The quenched fractions are used to describe the observed properties of local galaxies and they cannot be applied at all redshifts, which makes this method inapplicable to predictions of BBH mergers. For this reason, we have chosen not to model the quenching of star formation in massive galaxies with such considerations, but rather to quantify the impact of this discrepancy in our analyses (see a discussion in \S\ref{sub:caveats}).

\subsection{Models of cluster formation in a large scale cosmological simulation}
\label{sub:models}
Now that we have identified the galaxies within the \firebox\ simulation and have been able to extract their physical properties and their evolution over time, we want to describe the star clusters that would populate them. As individual star clusters are not resolved in \firebox, a cluster formation model is needed to get an estimate of the populations of massive star clusters that would exist within the galaxies of \firebox. In the following, we present three different sampling procedures used throughout this study for massive cluster formation in various galaxies located in a larger volume.

As \firebox\ uses a different value of gas density threshold to trigger star formation ($n\geq 300\mathrm{cm}^{-3}$ vs $n\geq 1000\mathrm{cm}^{-3}$ in the usual FIRE-2 simulations) and has a lower spatial resolution than the zoom-in simulations, we decided to not directly use the same cloud-to-cluster formation model to sample star cluster in \firebox. In order to take advantage of the unique dynamic range of \firebox\ and to maintain the same FIRE-2 physics models as in our previous studies \citepalias{grudic22,rodriguez23,bruel24}, we develop new models of massive star cluster formation that can be applied to this cosmological simulation.

We create three different simulation-based empirical models of massive star cluster formation based on the star clusters already sampled in our set of 6 zoom-in simulations (\texttt{m11q}, \texttt{m11i}, \texttt{m11e}, \texttt{m11h}, \texttt{m11d}, and \texttt{m12i}, with respective stellar masses $\Mstar=9.2\times 10^{8}$, $6.1\times 10^{8}$, $1.4\times 10^{9}$, $3.6\times 10^{9}$, $3.9\times 10^{9}$, and $6.7\times 10^{10}\ \Msun$ respectively, \citeauthor{elBadry17} \citeyear{elBadry17}, \citetalias{grudic22,bruel24}).
Here we enrich this sample of galaxies with three additional simulations from the FIRE-2 project: \texttt{m12r}, \texttt{m12c}, and \texttt{m12f} \citep{samuel19, kimmel19}. \texttt{m12r} is a medium-mass galaxy with $\Mstar\simeq 1.7\times10^{10}\Msun$, while \texttt{m12c} and \texttt{m12f} are both MW-like galaxies with $\Mstar\simeq5.8\times10^{10}\Msun$ and $\Mstar\simeq7.9\times10^{10}\Msun$ respectively. In all those simulated galaxies, we have used the \texttt{CloudPhinder} algorithm \citep{guszejnov19} to locate the giant molecular clouds (GMCs) in each snapshot and applied the cluster formation framework described in \citet{grudic21} to predict the population of star clusters that each GMC would produce.

The three models of massive star cluster formation proposed here each have a different level of complexity and are based on different assumptions, which makes it possible to highlight certain points of comparison when studying their predictions. In what follows, we describe in detail each of these models.

\subsubsection{Model `Gamma': Constant cluster formation efficiency}
\label{subsub:model1}
In this first approach, we make the straightforward assumption that, in a given galaxy, the star cluster formation rate is at any time a constant fraction of the total star formation rate:
\begin{equation}
    \dot{M}_{\mathrm{clusters}}[t] \equiv \Gamma \times \Psi[t] ,
\label{eq:GCrate1}
\end{equation}
where $\dot{M}_{\mathrm{clusters}}[t]$ is the formation rate of all star clusters at time $t$ in this specific galaxy, $\Gamma$ is its cluster formation efficiency assumed constant over time, and $\Psi[t]$ is its star formation rate at time $t$. 

In our sample of zoom-in simulations of galaxies, we observe that the most massive galaxies are more efficient at forming star clusters \citepalias[see][for more details]{bruel24}. To account for this trend, and for the possible variability of cluster formation in different galaxies with similar stellar masses, we proceed as follows:
\begin{itemize}
    \item based on the results obtained with individual simulations of galaxies, we fit the average cluster formation efficiency as a function of the galaxy stellar mass such that $\Gamma \simeq 0.043 + 0.019\times \mathrm{log}(\Mstar/10^9\Msun)$ with a standard deviation $\sigma \simeq 2.9\times 10^{-3}$.
    For a given \firebox\ galaxy, we use its stellar mass $M_{\star,90}$ to randomly draw one value of $\Gamma$ from the fit described above. Equation \ref{eq:GCrate1} now provides us with an estimate of the cluster formation rate in each galaxy.

    \item similarly, we observe in \citetalias{bruel24} and with the additional zoom-in cosmological simulations considered here that the distribution of cluster initial masses typically gets steeper with decreasing galaxy stellar mass. Consequently, we fit the average power-law index of the cluster initial mass function such that $\alpha \simeq -3.01 + 0.43\times\mathrm{log}(\Mstar/10^9\Msun)$ with a standard deviation $\sigma \simeq 0.11$. We use exactly the same approach in all three models to describe the dependence of the cluster mass function on the galaxy present-day stellar.

    \item finally, we use the combination of the cluster formation rate and the power-law slope of the distribution of cluster initial masses to sample individual cluster masses in all the snapshots of \firebox.
    In this sampling process, the minimum cluster mass possible is taken to be $10^3\ \Msun$ (same value used in \citetalias{grudic22} and \citetalias{bruel24}, and in the additional zoom-in simulations considered here).
    Since in this study, we are only interested in the production of BBHs in the most massive star clusters, and to maintain consistency with the cluster catalogues presented in \citetalias{grudic22,rodriguez23,bruel24}, we apply a mass threshold $M_\mathrm{cl} \geq 6\times 10^4\ \Msun$ to the collections of star clusters sampled in each galaxy. 
    This value roughly corresponds to the an initial number of particles ($\sim10^5$) which is the minimum required to ensure that the cluster relaxation timescale is always significantly longer than its dynamical timescale (a necessary condition in the evolution of star clusters with the code \cmc\ that we have used to integrate the clusters forward in time in \citetalias{rodriguez23} and \citetalias{bruel24}, see \S\ref{sub:grid} below for a brief description of this code).
\end{itemize}

\subsubsection{Model `EB18': Gas surface density}
\label{subsub:model2}
In this second part we present a sampling procedure similar to that described in \citet{elBadry18}. In this model, the formation efficiency of massive clusters is an increasing function of the gas surface density, which is precisely one of the features of the cluster formation model of \citet{grudic21} that has been used in our set of zoom-in simulations. Both theoretical models of star formation \citep[see e.g.][]{kruijssen12} and observation of star clusters in nearby galaxies \citep[see e.g.][]{portegies10} support this idea, with massive young star clusters being observed in high density environments. 
On the other hand, less dense environments have longer free-fall times and are less efficient at forming stars. This implies that these regions remain gas-rich and that the expulsion of this gas is more likely to unbound any cluster in formation.
Using high resolution simulations of collapsing clouds of gas (also evolved with the FIRE-2 model), \citet{grudic18} found that both the fraction of gas converted to stars, $\epsilon$, and the fraction of stars formed in bound clusters, $f_\mathrm{bound}$, can be expressed as functions of the physical properties of the giant molecular cloud (GMC), and primarily its surface density $\Sigma_\mathrm{GMC}=M_\mathrm{GMC}/R_\mathrm{GMC}^2$.
\citet{elBadry18} propose a functional form of this cluster formation efficiency as:
\begin{equation}
    \Gamma[t] \equiv \frac{\dot{M}_{\mathrm{massive}}[t]}{\Psi[t]} = \frac{\alpha_\Gamma}{1+(\Sigma_{\mathrm{GMCs}}[t]/\Sigma_\mathrm{crit})^{-\beta_\Gamma}} ,
\label{eq:GCrate2}
\end{equation}
where $\dot{M}_{\mathrm{massive}}[t]$ is the formation rate of massive clusters at time $t$, $\Psi[t]$ is the SFR at time $t$, and $\alpha_\Gamma$, $\beta_\Gamma$ and $\Sigma_\mathrm{crit}$ are constants that we estimate here from the massive star clusters already sampled in our zoom-in simulations of individual galaxies.

Insofar as our analyses indicated that the physical properties of the GMCs present in \firebox\ did not correspond exactly to what was expected from the zoom-in simulations, we decided not to take them into account here and to estimate the GMC surface density $\Sigma_\mathrm{GMC}$ following the process described in \citet{elBadry18}. Although this semi-analytical model was built and applied on a dark-model only merger tree, we decided to apply it to \firebox\ as is to serve as a point of comparison.
We describe here the different steps of this process.

Given a galaxy identified in \firebox, we compute at each snapshot, corresponding to the galaxy evolution time $t$, its star formation rate surface density as $\Sigma_{\Psi}[t]=\Psi[t]/\pi R_{\mathrm{d}}[t]^2$ where $R_{\mathrm{d}}[t]$ is the scale length of the gas disc at time $t$. This scale length is assumed to be a function of the halo specific angular momentum \citep{mo98} such that
\begin{equation}
    R_{\mathrm{d}}[t] = \frac{\lambda}{\sqrt{2}}R_{\mathrm{vir}}[t],
\end{equation}
where $\lambda$ is the halo spin parameter taken to be fixed at a typical value of $\lambda=0.035$
\citep{bullock01}.
Following \citet{elBadry18} the surface density of GMCs is then taken to be 
\begin{equation}
    \Sigma_{\mathrm{GMCs}} = 5\times 1.2\  10^3 \left(\frac{\Sigma_{\Psi}}{10\ \Msun\mathrm{kpc}^{-2}\mathrm{yr}^{-1}} \right) \Msun\mathrm{pc}^{-2} .
\end{equation}

The expression of the massive cluster formation efficiency in Equation \ref{eq:GCrate2} implies that very few massive clusters can be formed when the GMC surface density is low ($\Sigma_\mathrm{GMC}\ll\Sigma_\mathrm{crit}$), and it reaches the plateau value $\alpha_\Gamma$ when the GMC surface density is high ($\Sigma_\mathrm{GMC}\gg\Sigma_\mathrm{crit}$). Following \citet{elBadry18}, and supported by the values inferred from zoom-in simulations of collapsing clouds in \citet{grudic18}, we set $\Sigma_\mathrm{crit}=3000\ \Msun\mathrm{pc}^{-2}$ and $\beta_\Gamma=-1$.
We finally set $\alpha_\Gamma$ such that it is consistent with the total mass of massive clusters sampled in our set of zoom-in simulations used in \citetalias{bruel24}. This gives us a value $\alpha_\Gamma=0.06$. The strong difference between this value and the one proposed in \citet{elBadry18} of $\alpha_\Gamma=2.1\times 10^{-3}$ comes from the fact that they only consider massive clusters that survive to the present-day, while we aim to estimate the formation of all massive clusters, including those that get disrupted on short timescales. These clusters, which are also included in the zoom-in simulations, can contribute to the production of merging BBHs before their disruption, and even afterwards, if we consider the long time delays that BBHs can take to reach coalescence.

Once we have obtained the massive cluster formation rate from the galaxy SFR $\Psi[t]$ and the estimated formation efficiency $\Gamma[t]$, individual clusters masses are sampled from a power-law with a slope randomly drawn from the fit presented in \S\ref{subsub:model1} and with a minimum mass $M_\mathrm{min}=6\times10^4\ \Msun$.

\subsubsection{Model `SFRpeak': Extreme episodes of star formation}
\label{subsub:model3}
The third approach presented here relies on the idea that massive star clusters are not ordinarily formed in the most typical interstellar medium, but are a natural by-product of intense episodes of star formation. These episodes could themselves be related to environments with high gas density such as disks in high redshift galaxies \citep[see e.g.][]{kruijssen15} or to the outcome of galaxy interactions and mergers \citep[see e.g.][]{renaud16,renaud17,li17}.
To determine the epochs of cluster formation in each of the \firebox\ galaxies, we identify extreme episodes of star formation using the galaxy star formation history described in \S\ref{sub:FIREbox}.
In practice, we compute at each snapshot the ratio 
\begin{equation}
    \gamma[t] = \frac{\Psi[t]}{\overline{\Psi}_{\mathrm{1Gyr}}[t]} ,
\label{eq:GCrate3}
\end{equation}
where $\overline{\Psi}_{\mathrm{1Gyr}}[t]$ is the mean SFRD around time $t$. We use an arbitrary value of one Gyr to compute this mean SFR, as it shows reasonable results when applied to our set of zoom-in simulations (see Appendix). This ratio can then be used as a proxy for the cluster formation efficiency $\Gamma[t]$.

To predict the total mass of massive star clusters that should be sampled in each galaxy, we have used the populations of star clusters sampled in the zoom-in FIRE-2 simulations of individual galaxies and built a linear fit estimator. Similarly to the relation observed in a wide range of galaxies between the total mass of their GCs and their virial mass \citep[see e.g.][]{blakeslee97, harris17}, we relate the total mass of massive star clusters ($M_\mathrm{cl}\geq6\times 10^4\Msun$) that have ever formed in a given galaxy with its present-day stellar mass. With our set of zoom-in simulations, we obtain the relation $\mathrm{log}(\Sigma M_\mathrm{cl,massive})\simeq 5.97 + 2.01\times \mathrm{log}(\Mstar/10^9\Msun)$ with a standard deviation $\sigma \simeq 0.32$.

The combination of the estimated total mass of massive clusters to sample $M_\mathrm{cl,massive}$ with the locations of massive cluster formation indicated by $\gamma[t]$ provides us with the massive cluster formation rate of this third model. Here again, individual clusters masses are sampled from a power-law with a slope randomly drawn for each galaxy from the fit presented in \S\ref{subsub:model1} and with a minimum mass $M_\mathrm{min}=6\times10^4\ \Msun$.

\subsubsection{Sampling of other cluster properties}
Regardless of the sampling procedure for the number and masses of individual star clusters, we use the same following prescriptions to predict their other properties. 
To associate a value of metallicity to each cluster, we first determine the mean metallicity of its host galaxy at all snapshots following the same method used for the computation of the star formation histories (see \S\ref{sub:FIREbox}). Cluster metallicities are then interpolated from their formation times, with a scatter of 0.1 dex to account for a non-homogeneous metallicity in the considered galaxy at a given time. Although the dispersion of metallicities has been observed to be greater than this value for distant galaxies \citep[see e.g.][]{troncoso14} or lower in nearby galaxies \citep[see e.g.][]{williams21}, we find that this value gives is a good match to the metallicity dispersion for massive star clusters sampled in the individual zoom-in simulations.

Finally, the cluster half-mass radius is determined by sampling from a size-mass relation similar to that of \citet{grudic21}
\begin{equation}
    r_\mathrm{hm}=3\mathrm{pc}\left(\frac{M_\mathrm{cl}}{10^4\Msun} \right)^{1/3}\left(\frac{Z}{\Zsun}\right)^{1/10} ,
\end{equation}
with a log-normal scatter of $\pm0.4$ dex. 
Compared to the original formula presented in \citet{grudic21}, we have removed here the dependence of the clusters half-mass radii on the physical properties of their host GMCs $\left(\propto M_\mathrm{GMC}^{1/5}\Sigma_\mathrm{GMC}^{-1} \right)$, as we do not identify these structures individually.

\begin{figure*}
    \centering
    \begin{subfigure}{0.48\textwidth}
        \includegraphics{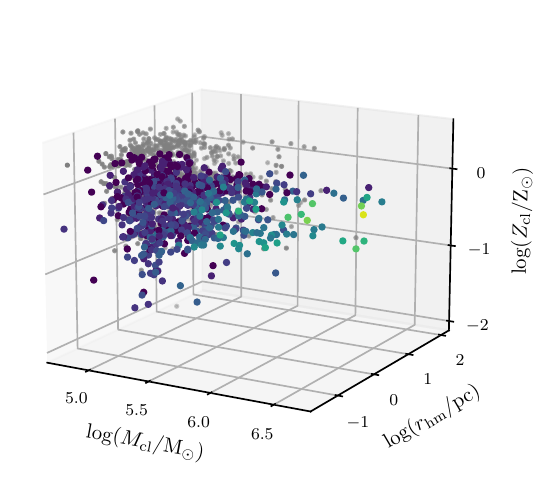}
    \end{subfigure}
    \hspace{1.5em}
    \begin{subfigure}{0.48\textwidth}
        \includegraphics{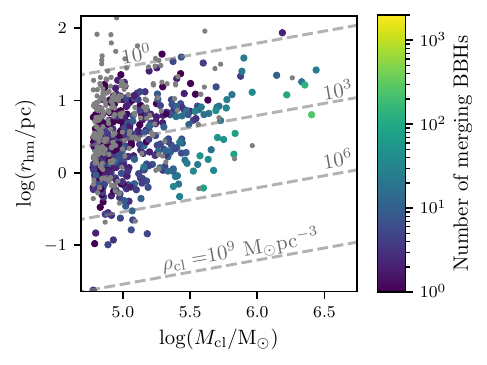}
        \vspace{1.8em}
    \end{subfigure}
    \caption{\textbf{Left panel:} 3D grid of the 1500 clusters integrated with \cmc, in parameter space ($M_\mathrm{cl}$, $r_\mathrm{hm}$, $Z_\mathrm{cl}$). Clusters in which no merging BBH are formed are indicated as grey dots, while the coloured dots represent the number of merging BBHs formed in each cluster.
    \textbf{Right panel:} 2D projection on the plane ($M_\mathrm{cl}$, $r_\mathrm{hm}$) for the subset of clusters with metallicities in the range $0.05\leq Z_\mathrm{cl}/\Zsun\leq0.5$ (623 star clusters out of the total 1500). The diagonal dashed grey lines represent lines of constant density $\rho_\mathrm{cl}\equiv M_\mathrm{cl}/(4/3) \pi r_\mathrm{hm}^3$. 
    }
    \label{fig:grid}
\end{figure*}

\subsection{BBH mergers in large populations of massive star clusters}
\label{sub:grid}

With almost a thousand galaxies identified in the simulation, and up to thousands of massive clusters sampled in each of them, it is certainly impractical to integrate them all with any code for cluster evolution \citep[with the possible exception of some recently developed rapid population synthesis codes for cluster evolution, such as the \texttt{RAPSTER} code][]{kritos24}. We develop a predictive model to estimate the BBH mergers formed in large populations of massive star clusters, by taking advantage of the 1500 massive clusters already integrated with the code \cmc\ in our previous studies \citepalias{rodriguez23, bruel24}. These clusters have been sampled in six different zoom-in cosmological simulations of individual galaxies from the FIRE-2 project, taking into account the impact of the galactic environment on their evolution through tidal fields and dynamical friction.

\cmc\ models collisional stellar dynamics in star clusters with an orbit averaging technique using a H\'enon Monte Carlo approach \citep{henon71}. It is based on the hypothesis that clusters with a sufficiently large number of particles ($\gtrsim 10^5$) evolve mainly by two-body encounters that can be modelled as a statistical process. This condition translates into the initial mass threshold $M_\mathrm{cl} \geq 6\times 10^4\ \Msun$ that we have applied to our clusters. Various processes relevant to the formation of BBHs are taken into account, including two-body relaxation \citep{joshi00}, three-body binary formation \citep{morscher15}, and direct integration of small-N resonant encounters with post-Newtonian corrections \citep{rodriguez18b}. The evolution of stars and stellar binaries in each cluster is modelled using the rapid population synthesis code \texttt{COSMIC} \citep{breivik20}.

From our collection of 1500 clusters evolved with \cmc\ we build a 3-dimensional grid in the parameter space ($M_\mathrm{cl}$, $r_\mathrm{hm}$, $Z_\mathrm{cl}$). We show this 3D grid in the left-hand panel of Figure \ref{fig:grid}, with the number of BBH mergers predicted by \cmc\ represented with different colours. The general trends that emerge from this collection of star clusters are that the most massive and densest clusters are the most likely to produce a large number of BBH mergers (see the right-hand panel of Figure \ref{fig:grid} for a projection on the ($M_\mathrm{cl}$, $r_\mathrm{hm}$) plane), and that the most metal-rich clusters are also the most inefficient. The latter result is due both to stellar evolution, which is particularly affected by metallicity through stellar winds and mass loss \citep[see e.g.][]{vink01,mapelli10,spera15,diCarlo20}, and to the fact that the majority of these clusters appear at very late times in their respective galaxies and have therefore not had the opportunity to evolve long enough for their BBHs to merge.

For every new cluster sampled in a given galaxy, we compute the Euclidean distance in log space with all the clusters in the grid as:
\begin{equation}
    d_\mathrm{i}^2 = \left(\mathrm{log}(\frac{M}{M_\mathrm{cl}})\right)^2 +\left(\mathrm{log}(\frac{r}{r_\mathrm{hm}})\right)^2
    + \left(\mathrm{log}(\frac{Z}{Z_\mathrm{cl}})\right)^2 ,
\label{eq:grid}
\end{equation}
where $M$, $r$, and $Z$ are respectively the mass, radius and metallicity of the new cluster considered.
To estimate the number of BBH mergers that this new cluster would produce, we first identify the 10 closest clusters in the grid according to the aforementioned distance (Equation \ref{eq:grid}).
We then compute the average of the number of BBH mergers predicted by \cmc\ for these 10 neighbouring grid clusters, weighted by the inverse of their distance to the new cluster. Finally, we sample the properties of these merging BBHs, namely the two component masses and the time elapsed between cluster formation and BBH merger, from the collection of merging BBHs produced in the 10 neighbouring clusters (again including inverse distance weighting).

In this form, the grid-matching method only takes into account the intrinsic properties of each star cluster sampled and does not consider the impact of the galactic environment on its evolution. And yet it has been shown that this impact is not negligible and plays an important part in determining the lifetime of star clusters \citep[see e.g.][]{rodriguez23}. We make here the assumption that the clusters already evolved with \cmc\ in our set of 6 zoom-in simulations \citepalias{bruel24} capture a wide range of tidal fields and that the selection of clusters in the grid is sufficiently random to ensure that the impact of these mechanisms on a large population of clusters is statistically respected.
Although we do not expect an exact match for every cluster sampled, especially since the effects of the galactic environment are not taken into account in this process of grid-matching, this method shows promising results when considering large populations of massive star clusters (see Appendix for a comparison with previous analyses from \citetalias{bruel24}).

\section{Results}
\label{sec:results}

\begin{figure*}
    \includegraphics{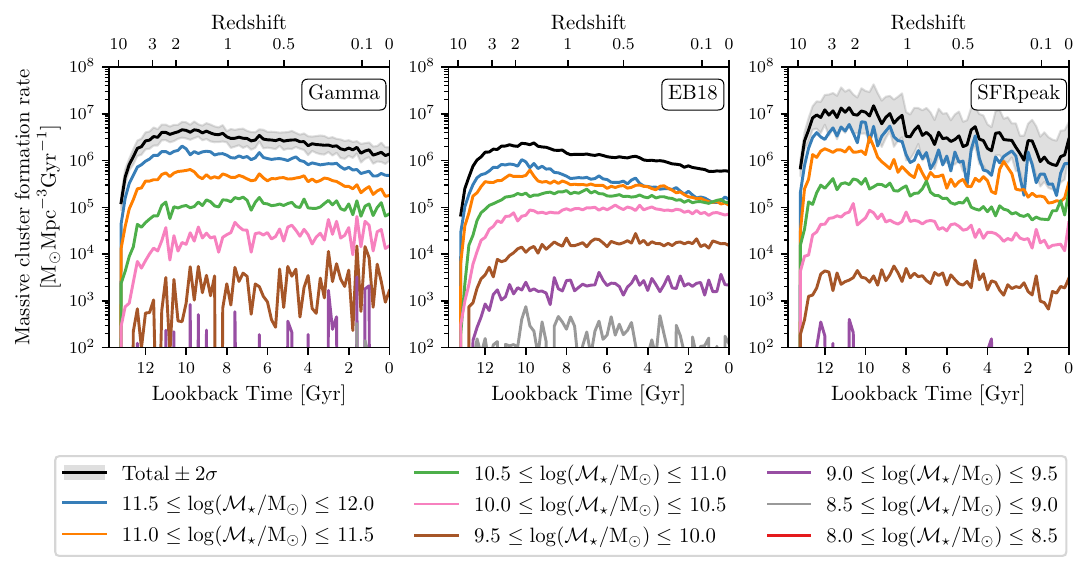}
    \caption{Formation rate of massive clusters ($M_\mathrm{cl}\geq6\ 10^4\Msun$) in \firebox\ galaxies using the cluster formation model `Gamma' (\textbf{left} panel), the `EB18' model (\textbf{middle} panel) and the `SFRpeak' model (\textbf{right} panel). 
    A re-weighting factor corresponding to the halo mass of the host galaxy (see \S\ref{sub:FIREbox}) is applied to each BBH obtained. Different colours indicate the contribution of galaxies grouped in different bins of stellar mass, with the same bins as those already used in Figure \ref{fig:SFRs in box}. The black solid line shows the mean total formation rate density obtained from 10 different random realisations, and the grey shaded area is the 90\% credible interval.}
    \label{fig:GCformrate}
\end{figure*}

\subsection{Massive star clusters sampled in \firebox}

The formation rates of massive clusters in all the galaxies identified in \firebox\ with the three different methods presented here are displayed in Figure \ref{fig:GCformrate}. All models predict a total formation rate density of massive star clusters in the range $10^5-10^6\ \Msun\mathrm{Mpc}^{-3}\mathrm{Gyr}^{-1}$ at $z=0$ and a peak of massive cluster formation at around redshift $z\sim2-3$, but with different values for this peak. The `SFRpeak', in particular, finds a formation rate density of massive clusters at high redshifts one order of magnitude higher than the two other models. It results in a decrease of this formation rate with a steeper, but also more stochastic, slope.
The `sawtooth' shape is a direct consequence of the small number of massive galaxies in \firebox. Indeed, the `SFRpeak' model places the epochs of massive cluster formation precisely in the most extreme episodes of star formation, and these massive galaxies each have important episodes of star formation at different epochs. This results in a formation rate density of massive star clusters that appears highly stochastic in such a reduced cosmological volume. In reality, we do not expect such strong stochasticity in the formation rate density of massive star clusters when considering a much larger cosmological volume.
The `Gamma' and `EB' models, however, find a much smoother formation rate over time. These features are expected from the very construction of these models, as they both follow the SFRD in each galaxy, and therefore the global SFRD in \firebox. 

In all three models the massive cluster formation rate is dominated at all redshifts by the more massive galaxies (blue and orange lines), even though there are far fewer of them (see Figure \ref{fig:SFRs in box}). It is clear that these predictions depend heavily on the physical processes that govern the growth of galaxies, and the omission of AGN feedback here plays a major role in the relative importance of the very massive galaxies ($\Mstar\geq10^{11}\ \Msun$). However, we do not expect that the inclusion of such feedback mechanisms would change the global picture of massive star clusters preferentially forming in massive galaxies ($\Mstar\geq10^{10}\ \Msun$).
On the other hand, it is also clearly apparent in the three panels that the low-mass galaxies ($\Mstar\leq10^9\ \Msun$) have a negligible contribution to the cosmological population of massive star clusters. Predictions for the intermediate-mass galaxies are fairly consistent across the three models.

This feature can be directly related to the excess of star formation that originates from the massive non-quiescent galaxies in \firebox\ (discussed above, see \S\ref{sub:FIREbox}).
We note here that, due to the number of massive non-quiescent galaxies in \firebox, the excess of star formation at low redshifts translates into an excess of galaxies with very high stellar masses (see Figure \ref{fig:HMF}). For the `Gamma' and `SFRpeak' models, which both use fits with galaxy stellar mass to estimate the populations of massive star clusters, this implies a local formation rate density of massive star clusters somewhat higher than the one predicted with the `EB18' model. However, this excess is associated with clusters with typically very high metallicities, which will therefore contribute very little to the overall production of merging BBHs (see the 3D-grid in the left-hand panel of Figure \ref{fig:grid} for the relation between cluster metallicity and BBH formation).

\subsection{Cosmological population of BBH mergers}
\label{sub:cosmo bbhs}

\begin{figure*}
    \includegraphics{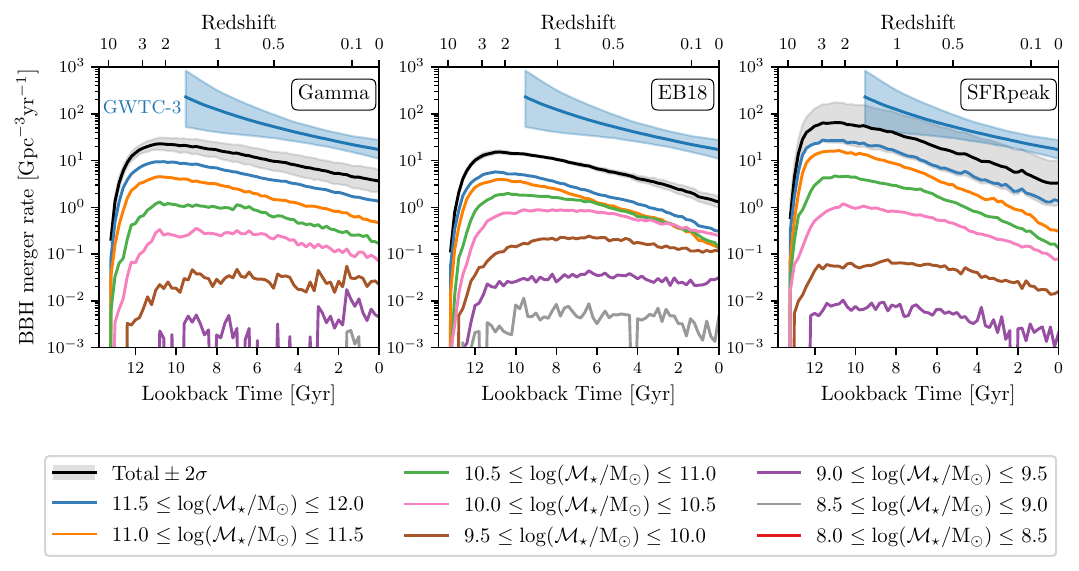}
    \caption{BBH merger rate density in \firebox\ estimated from the results of our cluster sampling algorithms (model `Gamma', `EB18', and `SFRpeak' presented from left to right respectively) combined with predictions of the 3D grid of massive clusters integrated with \cmc. A re-weighting factor corresponding to the halo mass of the host galaxy (see \S\ref{sub:FIREbox}) is applied to each BBH obtained. Different colours indicate the contribution of galaxies grouped in different bins of stellar mass. The black solid line shows the mean total merger rate density obtained from 10 different random realisations, and the grey shaded area is the 90\% credible interval. The blue shape shows the evolution of the BBH merger rate density with redshift as reported by the LVK Collaboration in \citet{gwtc3_pop}.}
    \label{fig:R in box}
\end{figure*}

We combine all the star clusters sampled with the 3D grid-matching technique described in \S\ref{sub:grid} to create populations of dynamically formed merging BBHs across cosmic time. The BBH merger rate densities obtained with the three models presented in \S\ref{sub:models} are shown in Figure \ref{fig:R in box}.
The three populations of BBH mergers give comparable estimates of the local BBH merger rate density from massive clusters in the range $\mathcal{R}_\mathrm{GCs}\sim1-10\ \mathrm{Gpc}^{-3}\mathrm{yr}^{-1}$.
These estimated values are in qualitative agreement with predictions from previous studies for the contribution of GCs to the astrophysical population of merging BBHs \citep[see e.g.][]{rodriguez18b,mapelli22}. 
\citet{fishbach23} infer a local BBH merger rate from GCs of $\mathcal{R}_\mathrm{GCs}(z=0)=10.9^{+16.8}_{-9.3}\ \mathrm{Gpc}^{-3}\mathrm{yr}^{-1}$ and increasing to $58.9^{+149.4}_{-46}\ \mathrm{Gpc}^{-3}\mathrm{yr}^{-1}$ at $z=1$. These predictions are more in line with those of the `SFRpeak' model.

From the detections of GWs, the LVK Collaboration infers an overall BBH merger rate density of $\mathcal{R}_\mathrm{BBHs}\sim17.9-44\ \mathrm{Gpc}^{-3}\mathrm{yr}^{-1}$ at a fiducial redshift $z=0.2$ \citep{gwtc3_pop}. The combination of the three cluster formation models `Gamma', `EB18', and `SFRpeak' with the grid-matching algorithm predict values of BBH merger rate density at $z=0.2$ of $4.8^{+3.1}_{-2.0}\ \mathrm{Gpc}^{-3}\mathrm{yr}^{-1}$, $2.8^{+0.7}_{-0.6}\ \mathrm{Gpc}^{-3}\mathrm{yr}^{-1}$ and $6.8^{17}_{-5}\ \mathrm{Gpc}^{-3}\mathrm{yr}^{-1}$ respectively. We emphasize here that the errors reported here do not represent the absolute uncertainties of our model, but rather the fluctuation inherent the stochasticity associated to the random processes taking place inside each of the three cluster formation models. 
The differences between these values and the one inferred by the LVK indicates that, though not dominant, the formation of merging BBHs in massive star clusters could very well represent a non negligible fraction of the astrophysical population.

It is particularly interesting to note that the three models have quite different slopes to describe the evolution of $\mathcal{R}_\mathrm{GCs}$ with redshift. In the hypothesis that we could accurately measure the evolution of the merger rate density with redshift and that we could separate BBH mergers coming from different formation channels (using, for example, the value of the chirp mass $\mathcal{M}_\mathrm{c}$, or values of the spin projections, see e.g. \citet{arcasedda20, antonelli23}), constraining in particular the redshift evolution of the dynamical BBH merger rate density would be particularly instructive for better understanding the formation of star clusters through cosmic time.

All three models of massive star cluster formation predict that the merger rate density of BBHs formed in massive star clusters is highly dominated by the most massive galaxies. This result seems to direct the identification of the host galaxies for the `dynamical' BBHs towards massive galaxies ($\Mstar\geq10^{10}\ \Msun$). Through a more detailed analysis, such feature could potentially be used in cosmological studies to assign astrophysically motivated probabilities to potential host galaxies for the most massive BBH mergers observed \citep[see e.g.][for different methods on how to use GW events as dark standard sirens for cosmology measurements]{mastrogiovanni23}.

\subsection{Properties of dynamically formed BBHs}
\label{sub:bbhs}

\begin{figure}
    \centering
    \includegraphics{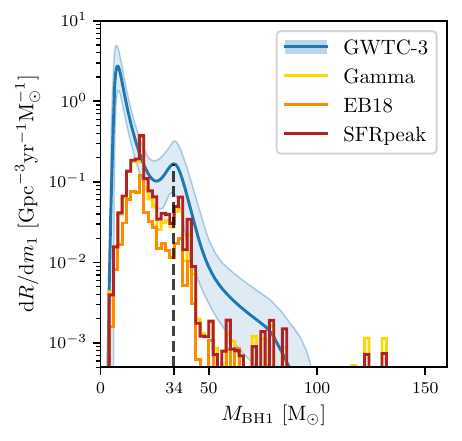}
    \caption{Astrophysical BBH primary mass distribution of all the local ($z\leq1$) BBH mergers sampled in \firebox. Different colours correspond to different massive star cluster formation models. We apply a re-weighting factor corresponding to the halo mass of the host galaxy (see \S\ref{sub:FIREbox}) to each BBH merger. 
    All the distributions are normalised to the predicted value for the BBH merger rate density at $z=0$. The blue curve shows the differential merger rate as a function of primary mass from GWTC-3 analyses \citep{gwtc3_pop}, with the shaded region showing the 90\% credible interval. 
    }
    \label{fig:mBH1 in box}
\end{figure}

We show in Figure \ref{fig:mBH1 in box} the distribution of BBH primary masses for the BBH mergers occurring at $z\leq1$ estimated from the results of our three cluster formation models (`Gamma', `EB18', and `SFRpeak') combined with predictions of the 3D grid of massive clusters already integrated forward in time with \cmc. The three distributions fall below the differential merger rate inferred by the LVK Collaboration with their fiducial \texttt{POWER LAW+PEAK} parametric mass model \citep[here in blue colour, taken from Figure 10 in][]{gwtc3_pop}, as is naturally expected from the merger rate densities estimated in this study (see \S\ref{sub:cosmo bbhs} and Figure \ref{fig:R in box}).
With all three cluster formation models, there is a distinctive feature at around $m_\mathrm{BH1}\sim35\ \Msun$ that is notably consistent with the location of the peak reported at $34^{2.6}_{-4.0}\ \Msun$ in \citet{gwtc3_pop} using the \texttt{POWER LAW+PEAK} model (indicated as the vertical dashed line in Figure \ref{fig:mBH1 in box}). We did not find any particular sampling effect responsible for the presence of this peak in our predictions. Further in-depth analyses of stellar evolution and cluster dynamics in the \cmc\ runs we use would be necessary to understand this feature in more detail.

It is also interesting to note that all three models predict a distribution of BBH primary masses that is clearly different from predictions of the isolated evolution channel obtained with rapid population synthesis codes \citep[see e.g.][]{vanson23}. The merging BBHs formed through dynamical interactions in massive star clusters can have masses that extend well above the pair-instability mass limit \citep[see e.g.][]{marchant19}, with an almost continuous distribution of primary masses from $\sim50$ to $\sim80\ \Msun$. The predictions of our astrophysical BBH primary masses in this mass range are remarkably consistent with the high-mass end of the \texttt{POWER LAW+PEAK} model, indicating a clear preference for the dynamical formation channel for such massive events.

Although our predictions on the physical properties of merging BBHs formed in massive star clusters do not provide an exact match to the astrophysical population inferred from GW observations, these features support the idea that different formation channels may populate different regions of the BBH mass distribution. Combined with BBHs formed through the isolated evolution of massive stellar binaries in the stable mass transfer channel \citep[see e.g.][]{vanson22b, vanson23}, and with BBHs formed in low and intermediate mass clusters \citep[see e.g.][]{torniamenti22}, both the predicted merger rate densities and the mass distribution could very well align with the astrophysical population inferred from GW observations.

We find no evolution of this BBH primary mass distribution with redshift for $m_\mathrm{BH1}\in[10-50]\ \Msun$, but a clear preference for the production of massive BBHs $m_\mathrm{BH1}>65\ \Msun$ in star clusters formed at $z\geq1.5$. This trend naturally arises from the fact that massive stars at low metallicity lose less mass in stellar winds, thus forming more massive compact objects, and low metallicity environments are preferably found at high redshifts.

Looking specifically at the galaxies in which these type of extreme mergers take place, we find that the majority of these events take place in the most massive galaxies (with present-day stellar masses $\Mstar\geq10^{10}\ \Msun$). Although a significant fraction of these massive galaxies should in reality be quiescent now, due to several feedback mechanisms not modelled in \firebox, the quiescent fraction is observed to be less important at higher redshifts \citep[see e.g.][]{fontana09}, and the absence of these feedback mechanisms in \firebox\ therefore has less impact on these predictions. 
This result indicates that the detection of such extreme BBH merger could most certainly translate into some strong constraints on the physical properties of its host galaxy. 
With the deployment of the next generation of GW interferometers, we may be able to identify the exact host galaxy of some of the nearest BBH mergers \citep[see e.g.][]{mo24}. The potential correlations observed between the physical properties of the host galaxies and those of the BBH mergers would make it possible to go even further in analysing the astrophysical origin of merging BBHs.

We note here that the three distributions are quite similar in shape because all the sampled BBHs come from the same 1500 clusters, which produce in total 7631 BBH mergers. 
With around a million massive star clusters sampled in total, each cluster in our 3D grid can be expected to be selected around 1000 times in average. 
In practice, we find that the average number of occurrences of each massive star cluster as nearest neighbour in the grid is indeed around 1000, with a handful of clusters selected more than 10000 times. We did not find any particular physical properties of these star clusters that are sampled most frequently.
This redundancy is then naturally reflected in the properties of the sampled BBHs. In practice, we find that the median number of occurrences of each BBH merger is around 100 for all three models of cluster formation. However, a small number of BBHs can be repeated several thousand times. Here again, we find no particular preference for the physical properties of the most frequently sampled BBHs.

Finally, we point out that a non-negligible number of BBH mergers sampled with all three cluster formation models in \firebox\ have primary masses well above $100\ \Msun$. With the three models of cluster formation `Gamma', `EB18', and `SFRpeak', we find respectively 2936, 1309, and 3106 BBH mergers taking place at redshifts $z<1$ in the volume of \firebox. These very extreme pairs of BHs are often the results of stellar mergers or hierarchical mergers that can only take place in very dense stellar environments. The lack of detection of BHs more massive than $100\ \Msun$ in LVK data seems to indicate that the models of stellar evolution and cluster dynamics presented here predict too massive objects \citep[see however][where higher harmonics are employed in the GW templates and, using a detection threshold $p_\mathrm{astro}>0.5$, they find 14 additional BBH mergers in LVK data, with the most massive one having $M_\mathrm{BH1}=300^{+60}_{-120}\ \Msun$]{wadekar2023}. Stronger observational constraints on the existence and detectability of such massive mergers will certainly emerge thanks to a greater number of GW observations and the contribution of new-generation detectors \citep[see e.g.][]{franciolini24}.

\subsection{Host galaxies of BBHs formed dynamically in massive star clusters}
\label{sub:host}

\begin{figure}
    \includegraphics{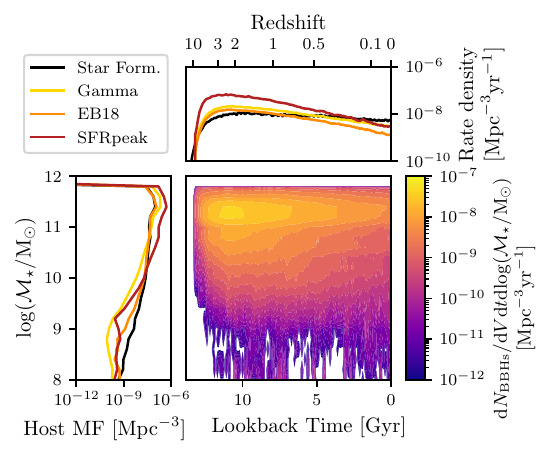}
    \caption{Merger times and galaxy present-day stellar masses of the BBH mergers in \firebox\ estimated from the `Gamma' cluster formation model (colormap). 
    The left and top subplots show the marginalised distributions obtained with the three different models of cluster formation (coloured lines), as well as the corresponding distributions for all star formation in the 979 galaxies identified inside \firebox\ (black solid line, with arbitrary normalisation).
    }
    \label{fig:host}
\end{figure}

We are now interested in the properties of the galaxies that contain the massive star clusters in which merging BBHs form dynamically. In Figure \ref{fig:host}, we show the distribution of galaxy present-day stellar masses and merger times of the BBH mergers sampled with the cluster formation `Gamma' in \firebox. These mergers preferentially originate from massive star clusters formed at high redshifts ($z\gtrsim2$) and in massive galaxies (with present-day stellar masses $\mathcal{M}_\star\gtrsim10^{11}\Msun$). We compare, for the three cluster formation models, the marginalised distributions of BBH merger over time (subplot on the left) and over galaxy masses (subplot on top). All three models qualitatively agree on the characteristic epochs and galaxy stellar masses for the production of BBH mergers in massive star clusters.

We also compare these distributions with all star formation in \firebox. Marginalised over galaxy masses this gives the total SFRD in the cosmological volume, already presented in Figure \ref{fig:SFRs in box}, and marginalised over time this gives the SFR-weighted density of galaxies at $z=0$. As mentioned before (\S\ref{sub:FIREbox}), most of star formation takes place around 10 billion years ago ($z\sim2$) in galaxies that are presently massive ($11\leq\mathrm{log}(\mathcal{M}_\star/\Msun)\leq12$).

Inspecting the marginalised distributions in more detail, we find an even more prominent production of BBH mergers at around $z\gtrsim3$ when compared with the normalised SFRD. The fact that relatively fewer massive clusters produce BBH mergers at low redshifts is a combined effect of higher metallicities and the necessary long delay times from cluster formation to the production of BBHs and all the way to their mergers due to GW emission alone. Comparing the galaxy mass functions, we find a similar preference for the highest present-day stellar masses, but also a clear relative dearth of low mass galaxies ($\mathcal{M}_\star\leq10^{10}\Msun$) contributing to the production of massive star clusters in which merging BBHs form. This difference arises from the fact that, although a number of stars form within them, there are practically no massive clusters in these low-mass galaxies.

The increasing sensitivity of current ground-based interferometers, as well as the future contribution of next-generation detectors in the coming decades, will certainly put strong constraints on the redshift evolution of the BBH merger rate density \citep[see e.g.][]{vanson22a}. This will be invaluable in helping us to better understand the mechanisms involved in the formation of these BBHs. Furthermore, these next-generation detectors will likely make it possible to localise some GW events in volumes small enough to allow for the identification of the host galaxies \citep[see e.g.][]{mo24}. We predict here that the majority of merging BBHs formed in massive star clusters originate from massive galaxies. This trend towards the formation of merging BBHs in massive galaxies is in agreement with previous predictions exploring the isolated evolutionary channel \citep[see e.g.][]{artale19a,artale19b}, although it has been shown to be strongly dependent on assumptions about star formation and metallicity distribution \citep[see e.g.][]{santoliquido22,srinivasan23}. On the other hand, \citet{srinivasan23} find that the stellar binary progenitors of detectable BBH mergers tend to form preferentially at lower redshifts ($z\lesssim1$) and in dwarf galaxies. Under the assumption that different formation channels of detectable BBH mergers typically take place in different environments and at different epochs, the identification of host galaxies for certain future GW events could shed new light on the question of the astrophysical origin of double compact objects.

\section{Discussion}
\label{sec:discussion}

\subsection{Caveats}
\label{sub:caveats}

One caveat to this study lies in the limits imposed by the cosmological volume simulation \firebox\ itself. Indeed, as described above (\S\ref{sub:FIREbox}) and in more detail in \citet{feldmann23}, there is a notable excess of star formation at $z>1$ in the simulation compared to observations. For the most part, this excess of star formation takes place in the most massive galaxies, whereas a large fraction of massive galaxies are observed to be quiescent. The lack of quenching of star formation in \firebox\ can be partly explained by the absence of any physical treatment of AGN feedback \citep[see][for more details]{feldmann23}.

With an excess of star formation at $z=0$ of a factor $\sim3$ in \firebox\ (see Figure \ref{fig:SFRs in box}), we can put an upper limit of the same factor on the over-prediction of the local BBH merger rate density originating from this overabundance of stars. However, since a certain fraction of local BBH mergers can be produced in massive star clusters formed at much higher redshifts where \firebox\ predicts a SFRD in good agreement with observations, the excess of star formation at low redshifts has in reality a more moderate impact. In practice, we find that, for the three models of cluster formation `Gamma', `EB18', and `SFRpeak' respectively, $52\%$, $50\%$ and $62\%$ of the BBH mergers taking place at redshifts $z\leq0.5$ originate from massive star clusters formed at $z>1$. Considering that, for the local BBH merger rate density, half of the BBHs are correctly produced in clusters formed at $z>1$ and the other half is overestimated by a factor 3 for clusters formed at $z<1$, this translates into a worst-case over-prediction factor of 2. 

This over-prediction contributes among other uncertainties associated with our cluster formation models and our grid-matching algorithm for predicting populations of dynamical BBH mergers. Further analyses need to be performed to quantify in more detail each of these uncertainties.

A second caveat lies precisely in the grid-matching algorithm, and in particular in the incomplete parameter space covered by the collection of our star clusters already integrated with the code \cmc. While, for most of the massive star clusters sampled, the closest neighbours in the grid have relatively similar physical properties (see e.g. Figure \ref{fig:distances} in the Appendix), a certain fraction of them can actually fall in a region of the parameter space that is not populated. This is notably the case for clusters with low metallicities, where our grid is only scarcely filled (see the 3D grid in Figure \ref{fig:grid}) This lack of low-metallicity clusters is largely due to the fact that \texttt{FIRE} cosmological simulations, including \texttt{m12i} which was used in \citetalias{grudic22, rodriguez23, bruel24}, typically have too little star formation at high redshifts \citep[see e.g.][]{wetzel23}.

This poses two problems. As \firebox\ uses the same physical model, we can expect that there will also be a lack of star formation at high redshifts, and therefore a lack of low-metallicity star clusters sampled in this cosmological volume simulation. Furthermore, the scarcity of low-metallicity clusters in our 3D grid means that our interpolation algorithm is unlikely to be very robust in this region of the parameter space.

To account for this second issue, we look at the impact of adding massive star clusters at low metallicities to the already existing 3D grid presented in \S\ref{sub:grid}. We select a set of 14 metal-poor clusters ($Z=0.01\ \Zsun$) from the total set of 148 systems presented in \citet{kremer20}. These clusters have all been evolved with the same code \cmc\, although with slightly different physical prescriptions (including the initial fraction of stellar binaries, which is taken as a constant value $f_\mathrm{b}=5\%$ in \citet{kremer20} while we have considered a mass-dependent binary fraction following \citet{vanHaaften13}).
After using the updated 3D grid to run the grid-matching algorithm on the populations of massive star clusters sampled with our three formation models, we find no major difference in the predictions already presented in Section \ref{sec:results}.

This does not come as a surprise, as the addition of only 14 star clusters to a grid containing already 1500 is not expected to have a significant impact. To quantify the impact of this empty region of the parameter space, more clusters could potentially be run with \cmc. Different codes for star cluster evolution could also be used for a more global comparison. We leave the exploration of these aspects to future studies.

\subsection{Comparison with other studies}
\label{sub:comparison}

Our three models of massive cluster formation predict a total formation rate density of massive star clusters in the range $10^5-10^6\ \Msun\mathrm{Mpc}^{-3}\mathrm{Gyr}^{-1}$ at $z=0$ and a peak of massive cluster formation at around $z\sim2-3$ (see Figure \ref{fig:GCformrate}). These high values of the massive cluster formation rate at $z>2$ are in qualitative agreement with observations of star-forming galaxies at high redshifts \citep[see e.g.][]{shapiro10}. 
Integrated over time, we find a value of the cluster mass formed per unit volume of $\rho=3.2,\ 1.7,\ \mathrm{and}\ 7.3\times10^7\ \Msun\mathrm{Mpc}^{-3}$ with our three cluster formation models `Gamma', `EB18', and `SFRpeak' respectively. As a comparison, by constructing their own model of GC formation with a present-day mass density of GCs in the Universe consistent with its empirical value and a current mass function consistent with the observed distribution of galactic GCs, \citet{antonini20} find $\rho=2.4\times10^7\ \Msun\mathrm{Mpc}^{-3}$ with an estimated uncertainty of a factor of 2. 
As a comparison, \citet{antonini20} construct their own model of GC formation using a present-day mass density of GCs in the Universe consistent with its empirical value, and a present-day mass function consistent with the distribution of observed galactic GCs. They find a value of the cluster mass formed per unit volume $\rho=2.4\times10^7\ \Msun\mathrm{Mpc}^{-3}$ with an estimated uncertainty of a factor of 2. This falls reasonably within the predictions of our models.

Using the catalogue of GW sources GWTC-3 \citep{gwtc3_pop}, and assuming that the GC formation rate number density follows a Madau-like \citep{madau14} functional form, \citet{fishbach23} infer a value for the GC formation rate at redshift 2 of $\sim2\times10^6\ \Msun\mathrm{Mpc}^{-3}\mathrm{Gyr}^{-1}$. This value is in good agreement with the predictions from the two models `Gamma' and `EB18'. The cluster formation model `SFRpeak' predicts overall a more important formation rate of massive star clusters. This feature could very well be explained by the fact that it uses scaling relations with galaxy present-day stellar masses, and the most massive galaxies are found to be in excess inside \firebox. A more realistic cosmological volume population or a post-processing treatment for AGN feedback in \firebox\ could potentially resolve this issue and produce more consistent predictions between our different cluster formation models.

\section{Conclusions}
\label{sec:conclusions}

In this work we have modelled the formation of massive star clusters in the cosmological volume simulations \firebox\ (part of the FIRE-2 project). We have used three different models of massive cluster formation empirically built from results of previous studies \citep{grudic22,rodriguez23,bruel24} based on a cloud-to-cluster formation framework applied to the GMCs identified in a set of cosmological zoom-in simulations of individual galaxies. Extending on the work presented in these studies, we have assembled all the massive star clusters already integrated with the Monte Carlo code \texttt{CMC} in a 3-dimensional grid in the parameter space ($M_\mathrm{cl}$, $r_\mathrm{hm}$, $Z_\mathrm{cl}$) and developed an algorithm to predict the population of merging BBHs that would be produced by any large population of massive clusters. Combined with our different models of cluster formation applied to close to a thousand galaxies identified in \firebox, this has enabled us to create populations of merging BBHs that would have formed in these realistic environments.

We have studied the properties of these BBH mergers and of their host galaxies. We have found distinct features that could prove decisive in better understanding the formation history of massive star clusters and their contribution to the astrophysical population of BBH mergers in the local Universe.

\begin{enumerate}
\item Different assumptions on the mechanisms driving the production of massive star clusters in various galaxies result in different predictions for their formation rate over cosmic time. In particular, assuming that the formation of massive star clusters occurs preferentially during episodes of intense star formation results in a formation rate density of massive star clusters with a larger peak at early times ($z<2$) and with a steeper decrease up to the present-day. In contrast, assuming that their formation follows the global star formation rate results in smoother evolutions over time.
\item Over different models of massive cluster formation, we have predicted a consistent value of the local BBH merger rate density in the range $\mathcal{R}_\mathrm{GCs}\sim1-10\ \mathrm{Gpc}^{-3}\mathrm{yr}^{-1}$. Although below the value inferred by the LVK Collaboration from GW signals, it is a clear indication that this formation channel may contribute to a significant fraction of the list of detected BBH mergers. Different models of cluster formation model predict different vales for the local BBH merger rate density, but also for its evolution with redshift.
\item In agreement with previous studies, we have found that the physical properties of BBHs formed in massive star clusters hold very distinct features. The possibility of stellar mergers and hierarchical BBH mergers in such dense environments allows for the formation of very massive BHs. These extreme systems, which are challenging to explain via different formation channels, bear the imprint of their dynamical formation in their unique physical properties.
\item The massive clusters that produce BBHs merging in the local Universe are found to exist in galaxies with similar properties as those driving star formation at a cosmological scale: they typically form at $z\gtrsim2$ in galaxies with present-day stellar masses in the range $11\leq\mathrm{log}(\mathcal{M}_\star/\Msun)\leq12$. While the peaks of these two distributions lie in the same region, we have also found a clear relative dearth of massive clusters producing BBH mergers in low mass galaxies ($\mathcal{M}_\star\leq10^9\ \Msun$). It is worth noting that these results run counter to some predictions of the isolated evolution channel. This indicates that identifying the host galaxies of BBH mergers could provide valuable constraints on their formation channels and on the branching fraction between these channels.
\end{enumerate}

This study represents the first attempt to estimate the populations of merging BBHs formed in massive star clusters in the cosmological volume simulation \firebox. These results highlight the importance of modelling the formation of star clusters for predicting their contribution to the astrophysical population of local BBH mergers but also to the redshift evolution of the BBH merger rate density. The ever-growing list of observed BBH mergers, the increasing sensitivities of the current GW interferometers, and the future advent of the third generation detectors will provide valuable information on the origin of the detected BBH mergers, on the redshift evolution of the merger rate density, and therefore on the properties of the massive star clusters that produce them.

\begin{acknowledgements}
    Tristan Bruel is supported by ERC Starting Grant No.~945155--GWmining, Cariplo Foundation Grant No.~2021-0555, MUR PRIN Grant No.~2022-Z9X4XS, MUR Grant ``Progetto Dipartimenti di Eccellenza 2023-2027'' (BiCoQ), and the ICSC National Research Centre funded by NextGenerationEU.
    Tristan Bruel and Astrid Lamberts are supported by the ANR COSMERGE project, grant ANR-20-CE31-001 of the French Agence Nationale de la Recherche.  Carl L.~Rodriguez acknowledges support from NSF Grant AST-2310362, NASA ATP Grant 80NSSC22K0722, a Alfred P.~Sloan Research Fellowship, and a David and Lucile Packard Foundation Fellowship.  This work was supported by the ‘Programme National des Hautes Energies’ (PNHE) of CNRS/INSU co-funded by CEA and CNES’ and the authors acknowledge HPC ressources from ‘Mesocentre SIGAMM’ hosted by Observatoire de la C\^ote d’Azur. This work made use of infrastructure services provided by the Science IT team of the University of Zurich (\url{www.s3it.uzh.ch}). The \firebox\ simulation was supported in part by computing allocations at the Swiss National Supercomputing Centre (CSCS) under project IDs s697, s698, and uzh18. Support for MYG was provided by NASA through the NASA Hubble Fellowship grant \#HST-HF2-51479 awarded  by  the  Space  Telescope  Science  Institute,  which  is  operated  by  the   Association  of  Universities  for  Research  in  Astronomy,  Inc.,  for  NASA,  under  contract NAS5-26555. 
\end{acknowledgements}

\bibliographystyle{aa} 
\bibliography{references}

\begin{figure*}
    \centering
    \includegraphics{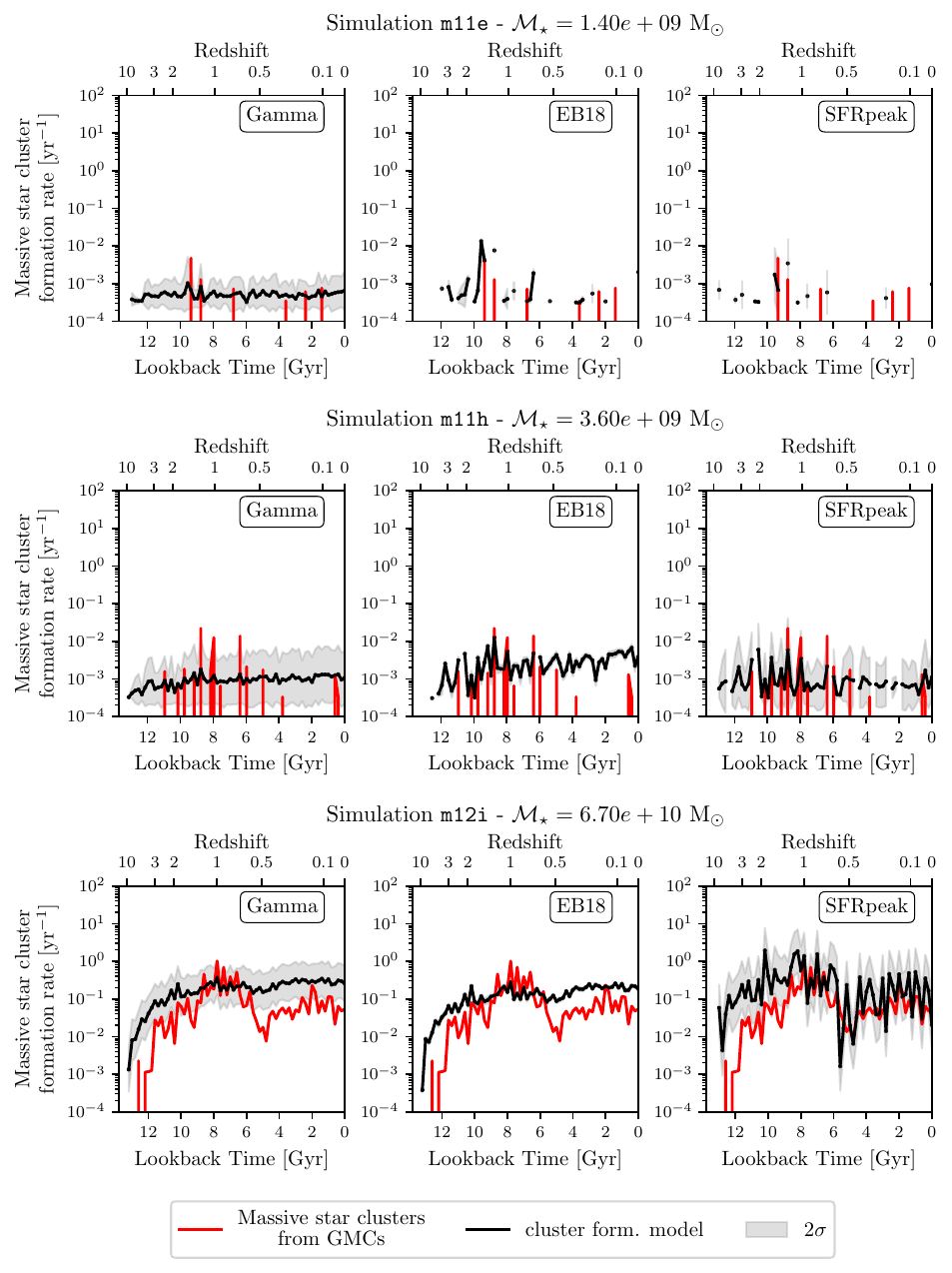}
    \caption{Massive star cluster formation rate estimated in the cosmological zoom-in simulations \texttt{m11e}, \texttt{m11d}, and \texttt{m12i} from top to bottom by sampling massive star clusters from individual GMCs \citepalias[red, see][]{bruel24}, and from the results of our cluster sampling algorithms (models `Gamma', `EB18', and `SFRpeak' respectively presented from \textbf{left} to \textbf{right}). For each model, the process is repeated 100 times and we show the mean formation rate (black solid line) as well as the 90\% credible interval (grey shaded area).}
    \label{fig:clusters_m12i}
\end{figure*}

\begin{figure*}
    \centering
    \includegraphics{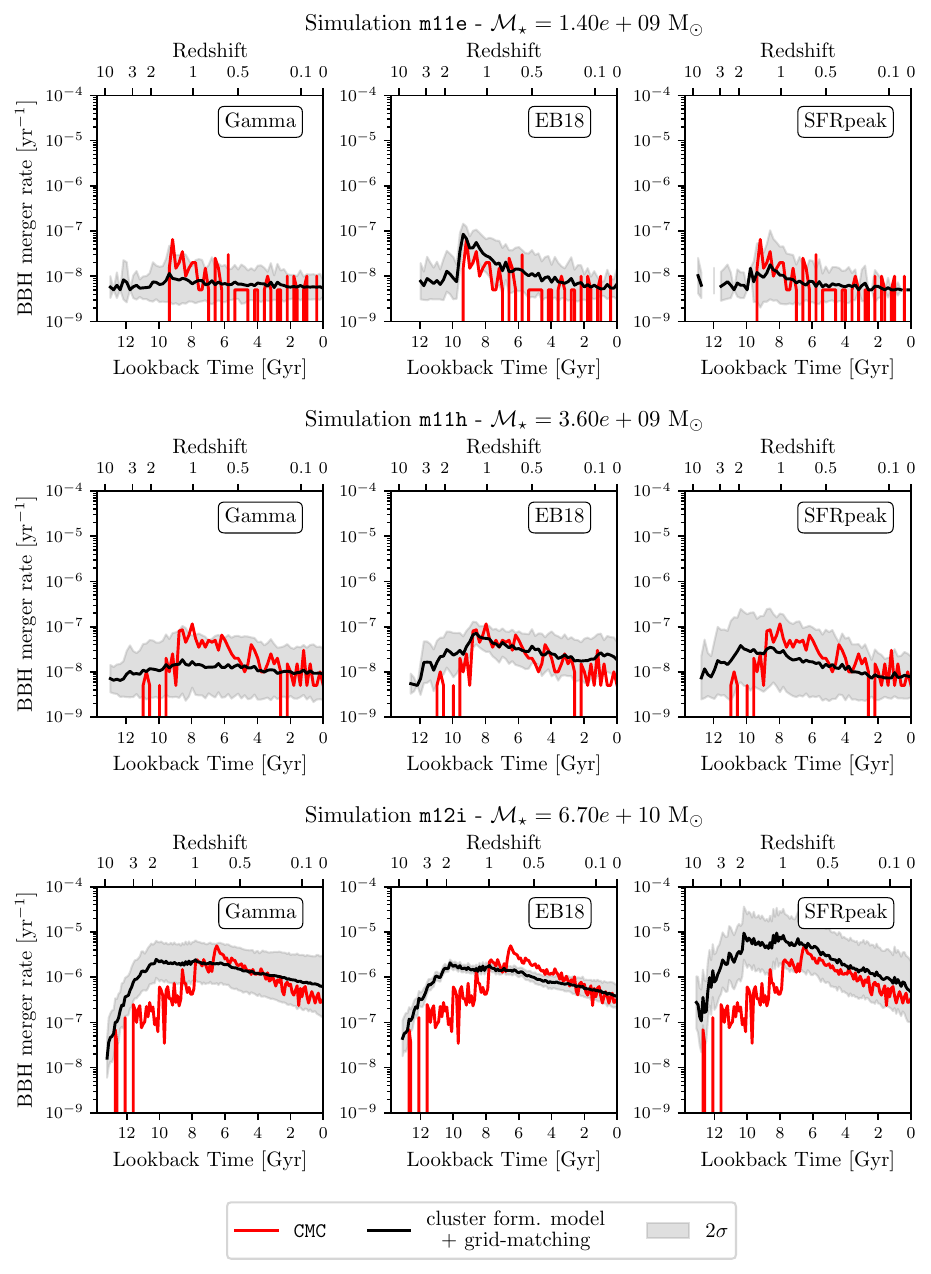}
    \caption{BBH merger rate computed in the cosmological zoom-in simulations \texttt{m11e}, \texttt{m11d}, and \texttt{m12i} from top to bottom by sampling massive star clusters from individual GMCs and integrating them forward in time with the code \cmc\ \citepalias[red, see][]{bruel24}, and from the results of our cluster sampling algorithms (models `Gamma', `EB18', and `SFRpeak' respectively presented from \textbf{left} to \textbf{right}) combined with predictions of the 3D grid. For each model, the process is repeated 100 times and we show the mean BBH merger rate (black solid line) as well as the 90\% credible interval (grey shaded area).}
    \label{fig:bbh_m12i}
\end{figure*}

\appendix

\renewcommand\thefigure{\arabic{figure}}
\setcounter{figure}{9}

\section*{Quantitative testing of the three sampling methods of massive star clusters combined with the grid-matching algorithm for predicting BBH mergers}
\label{appendix:verification}

\begin{figure}
    \centering
    \includegraphics{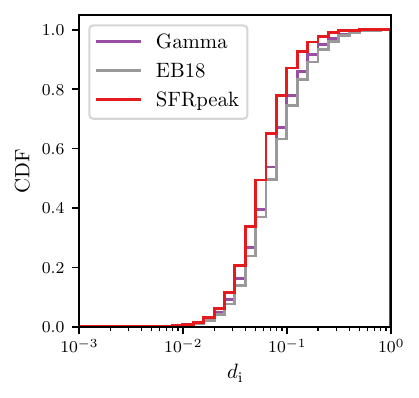}
    \caption{Cumulative distribution of distances $d_\mathrm{i}$ obtained after sampling massive clusters in the simulation \texttt{m12i} with the three models `Gamma', `EB18', and `SFRpeak', and applying the ~grid-matching scheme.}
    \label{fig:distances}
\end{figure}

As a consistency test, we apply the three different models of massive star cluster formation presented in \S\ref{sub:models} to the simulated galaxies \texttt{m11e}, \texttt{m11h}, and \texttt{m12i} \citepalias[see][for further analysis on the GMCs, massive star clusters, and BBH mergers in this simulated galaxy]{grudic22, rodriguez23, bruel24}. The formation rates of massive star clusters obtained are shown in Figure \ref{fig:clusters_m12i}. As there is some uncertainty due to random draws in the cluster sampling models, in particular for the `Gamma' and `SFRpeak' models, we repeat the whole process 100 times to obtain different estimates of the star cluster populations in each simulated galaxy for each of the cluster models. There is clearly no exact match between the clusters sampled for the three cluster formation models and the ones obtained from identified GMCs, particularly at low redshifts. This is due to the fact that, while star formation still occurs at low redshifts in the simulation \texttt{m12i}, few massive and dense GMCs are found there, which translates into few massive star clusters. The `SFRpeak' model, locating the epochs of massive cluster formation during extreme episodes of star formation only, provides a better match to the cluster formation rate at these low redshifts, but also over-predicts their formation at higher redshifts.

We then apply the grid-matching algorithm presented in \S\ref{sub:grid} to estimate populations of dynamically formed merging BBHs. 
Figure \ref{fig:bbh_m12i} shows the results of this test in the form of a BBH merger rate and compares with the merger rate obtained with massive star clusters sampled from individual GMCs and integrated forward in time with \cmc \citepalias{bruel24}. In this example, all models struggle to match the BBH merger rate at high redshifts, but they recover the local ($z\leq1$) BBH merger rate with a fairly good accuracy. This result is quite encouraging, as the three cluster formation models were built using the star clusters from our set of zoom-in simulations, and without taking their BBH mergers into account. We thus find that the combination of approximate models of cluster formation and BBH production developed here remains consistent with the more detailed study carried out on simulations of individual galaxies.

A more complete analysis shows that the discrepancy observed at redshifts $z>1$ is not mainly due to the grid matching method, but rather to differences in the number and properties of massive clusters predicted to exist at these epochs.
Indeed, examining in detail the predictions of the three proposed cluster formation models, it appears that they all tend to overestimate the formation of massive clusters at early times compared with clusters sampled directly from the GMCs. This is linked to the fact that our three models use the SFRD of a galaxy to estimate the formation of massive clusters within it, whereas the appearance of massive and dense GMCs, which were precisely used as the birthplaces of clusters in the previous study, is not necessarily correlated with this SFRD. This eventually results in a BBH merger rate always overestimated at these high redshifts. Since we aim to compare our predictions with LVK observations of BBH mergers which, with current detector sensitivities, are found at redshifts $z\leq1$, these discrepancies at high redshifts can be set aside for the moment.

To measure the precision of the grid-matching method, we store the list of distances to the closest neighbour in the grid computed for all the clusters sampled for each of the three cluster formation models (one iteration of cluster sampling in the simulated galaxy \texttt{m12i}). We show the cumulative distributions of the obtained distances in Figure \ref{fig:distances}. In this example, all the clusters sampled using the three models have a nearest neighbour in the 3D grid at a distance of less than 1, and more than $81\%$ of them with $d_\mathrm{i}\geq0.15$. This last value corresponds to a cluster whose physical properties are typically 1.2 times higher (or lower) than those of its nearest neighbour.

\end{document}